\documentclass[aps,prl,twocolumn,showpacs,superscriptaddress,superscriptaddress]{revtex4-1}  
\usepackage{graphicx,color}  
\usepackage{dcolumn}   
\usepackage{bm}        
\usepackage{amssymb}   
\usepackage{ulem}
\usepackage[section]{placeins}
\usepackage{float}
\usepackage{amsmath}
\usepackage{epstopdf}

\begin{document}

\title{Scaling Description of Non-Local Rheology}

\author{Thomas Gueudr\'e}
\affiliation{DISAT, Politecnico Corso Duca degli Abruzzi, I-10129 Torino - Italy}
\author{Jie Lin}
\affiliation{Center for Soft Matter Research, Department of Physics, New York University, New York, NY 10003}
\author{Alberto Rosso}
\affiliation{LPTMS, CNRS, Univ. Paris-Sud, Universit\'e Paris-Saclay, 91405 Orsay, France}
\author{Matthieu Wyart}
\affiliation{Institute of  Physics, Ecole Polytechnique Federale de Lausanne (EPFL), CH-1015 Lausanne, Switzerland}

\date{\today}
\begin{abstract}
Non-locality is crucial to understand the plastic flow of an amorphous material, and has been successfully described by the \textit{fluidity}, along with a cooperativity length scale $\xi$. We demonstrate, by applying the scaling hypothesis to the yielding transition, that non-local effects in non-uniform stress configurations can be explained within the framework of critical phenomena. From the scaling description, scaling relations between different exponents are derived, and collapses of strain rate profiles are made both in shear driven and pressure driven flow. We find that the cooperative length in non-local flow is governed by the same correlation length in finite dimensional homogeneous flow, excluding the mean field exponents. We also show that non-locality also affects the finite size scaling of the yield stress, especially the large finite size effects observed in pressure driven flow. Our theoretical results are nicely verified by the elasto-plastic model, and experimental data.

\end{abstract}

\maketitle

Amorphous materials including emulsions, foams and glasses are important for various industrial processes. 
They are yield stress materials, which transition from a solid to a liquid state as the shear stress passes some threshold $\Sigma_c$. In many instances, this transition appears to be a collective phenomenon. In the liquid phase, the dynamics becomes correlated on a length scale $\xi\sim (\Sigma-\Sigma_c)^{-\nu}$ near $\Sigma_c$ \cite{Picard2005,Lemaitre09,Karmakar10,Lin14,Liu16}, and the flow curve is singular: 
\begin{align}
\dot{\gamma} \sim \left( \Sigma -\Sigma_c \right)^{\beta},
\label{HerschelBulkley}
\end{align}
where $\dot{\gamma}$ is the strain rate and $1/\beta$ is the Herschel-Bulkley (HB) exponent \cite{Bonn15}. However, the most obvious signature of collective behavior  are non-local effects \cite{nichol2010flow,komatsu2001creep,reddy2011evidence,Goyon08,Bocquet09,Chaudhuri12,Jop12,Bouzid13,bouzid2015non,Kamrin12,nicolas2013mesoscopic}: as $\Sigma_c$ is approached, the presence of a boundary (such as a wall) or of inhomogeneities in the applied stress affects flow  on a growing length scale, often fitted as $\ell(\Sigma)\sim (|\Sigma-\Sigma_c|)^{-\tilde \nu}$. Such behavior is particularly important  for micro and nano-fluidic devices \cite{Goyon08}.
At a microscopic level, there is a growing consensus that flow consists of local rearrangements, the so-called shear transformations \cite{Argon79,Falk98}, which are coupled to each other by long-range elastic interactions \cite{Picard04}.  Indeed elasto-plastic models (cellular automata) \cite{Baret02,Picard2005} incorporating these ingredients can reproduce the mentioned  collective  effects.

However, which theoretical framework can quantitatively explain these behaviors remains debated.  On the one hand, a popular approach has focused on non-local effects, and introduced  the concept of {\it fluidity} $f(\vec x)$, namely the rate of plasticity at position $\vec x$. Elastic coupling between plastic events is then described by a phenomenological laplacian equation \cite{Goyon08}:
\begin{equation}
\nabla^2 f(\vec x)=(f(\vec x)-f_b(\Sigma))/\ell(\Sigma)^2
\label{standard_fluidity}
\end{equation}
where $f_b=\dot{\gamma}/\Sigma$ is the bulk value of the fluidity as governed by Eq.(\ref{HerschelBulkley}). 
Eq.(\ref{standard_fluidity}) was later justified \cite{Bocquet09,Chaudhuri12} in a mean-field framework for which  $\tilde \nu_{MF}=1/2$ and $\beta_{MF}=2$.
Although experiments with concentrated emulsion\cite{Jop12} and elasto-plastic simulations\cite{Nicolas13} fitted flow profiles with $\tilde\nu=0.5$, obvious deviations remained. A constant $\ell$, independent of $\Sigma$, has even been proposed for emulsion in micro-channels \cite{Goyon08,Goyon10}.

On the other hand, we have recently introduced a scaling description of homogeneous flows  \cite{Lin14} that  predicts  several relations between exponents, supported by recent numerics \cite{Salerno13,Lin14,Liu16}. In this framework,  exponents can be proven \cite{Lin14a,Lin16} to be non mean-field, in agreement with the observations that $\nu_{2d}\approx 1.16$ and $\nu_{3d}\approx 0.72$ \cite{Salerno13,Lin14}.  These finding do not necessarily contradict the fluidity theory, as it could be that two length scales exist and $\tilde \nu\neq \nu$. An alternative possibility is that the fluidity approach is not  exact. If so, can an exact phenomenological description be proposed, and used to relate bulk properties and non-local effects?

\begin{figure}[hbt!]
\centering\includegraphics[width=.45\textwidth]{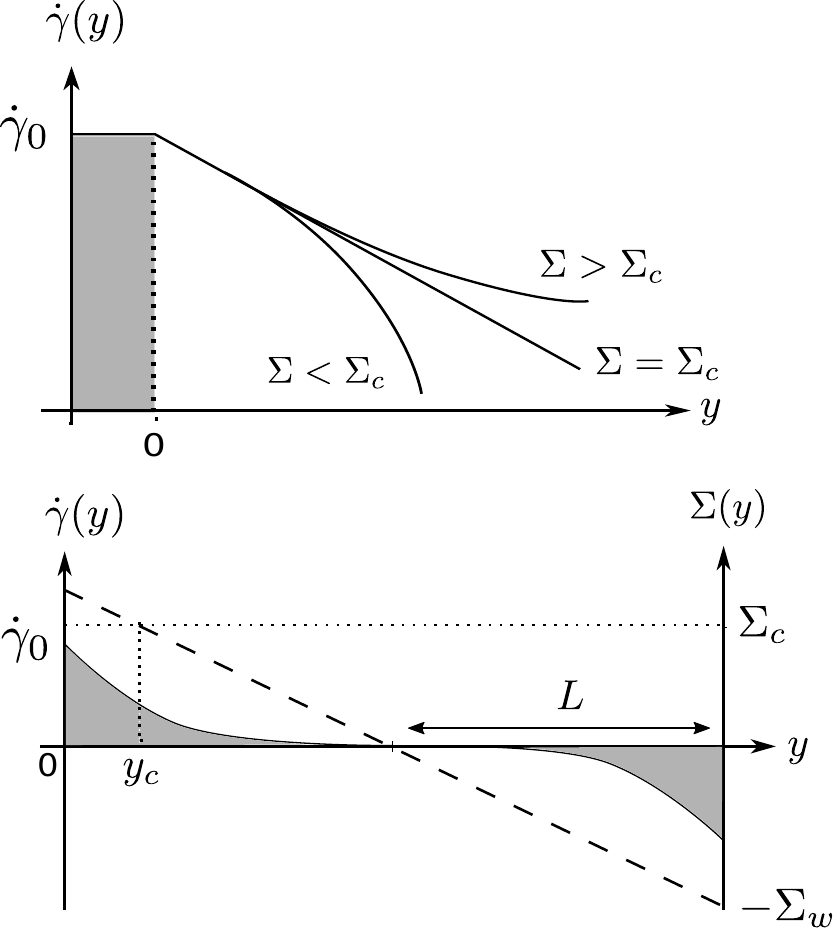} 
   \caption{Illustration: in the presence of a permenant shear band at the boundary, the strain rate in the bulk with a uniform stress $\Sigma$ decays to zero if $\Sigma<\Sigma_c$, or to the homogeneous value $\dot{\gamma}\sim (\Sigma-\Sigma_c)^{\beta}$ if $\Sigma\gtrsim\Sigma_c$. Right at $\Sigma_c$, the strain rate decays as a power law $\dot{\gamma}(y)\sim y^{-\alpha}$, which is shown as the straight blue line in the log-log plot. Sketch describing the pressure driven flow. The dashed line is the background stress $\Sigma(y)$, starting at the wall at $\Sigma_w$. The dotted line is the critical stress $\Sigma_c$. The curve delimiting the grey area is a typical activity profile $\dot{\gamma}(y)$, penetrating into the solid phase $2L-y_c>y>y_c$}\label{illus1}
\end{figure}

In this letter we build a scaling description of non-local effects. We first focus on the case where a shear band at fixed $\dot\gamma$ is connected to a system at fixed stress $\Sigma$ shown in Fig.~\ref{illus1} (top). We derive a scaling relation for the flow profile which can be used  to extract $\tilde \nu$ from data collapse. We test this procedure in elasto-plastic models and show that $\nu\approx \tilde \nu$ with a very good accuracy, supporting that a single length scale characterizes the yielding transition. This procedure definitely excludes the mean-field value $\tilde \nu=1/2$. Our analysis shows that Eq.(\ref{standard_fluidity}) does not satisfy basic requirements imposed by critical behavior near surfaces, and may thus lead to improper extraction of $\ell(\Sigma)$. We propose an alternative local differential equation for the flow profile for varying stress, 
and discuss its limitations. Finally, we extend our analysis to pressure driven flows where our predictions are satisfyingly compared to experiments and numerical models, and explain the previous observation that finite size effects are  weaker in this geometry.

\hfill

{\it Effect of a shear band:} Experimentally it is observed that if the applied stress (or threshold stress) varies in space such that only a portion of the material is above threshold, a shear band appears that creates a mechanical noise that spreads in the system.  As a result, flow occurs everywhere in the material, even below threshold \cite{PhysRevLett.104.078302,nichol2010flow,komatsu2001creep,reddy2011evidence,PhysRevLett.104.078302}, a phenomenon akin to the creep motion of ferromagnetic domain walls \cite{Lemerle98,Kolton06,Ferrero16}.
The simplest configuration to study such effects is shown in Fig.~\ref{illus1} (top), where a shear band with strain rate  of order one is imposed for $y<0$, while for $y>0$ the  stress $\Sigma$ is constant. 

In this geometry, we hypothesize (as commonly assumed near critical points \cite{henkel2008non}) that the only relevant length scale is $\ell(\Sigma)\sim |\Sigma-\Sigma_c|^{-\tilde \nu}$. The fluidity at the vicinity of $\Sigma_c$ can then only depend on $y$ via the ratio $y/\ell(\Sigma)$, and can thus be written as:
\begin{align}
f(y)\equiv \dot{\gamma}(y)/\Sigma & = |\Sigma-\Sigma_c|^{c_0}  \mathcal{F}_{\pm}( y |\Sigma-\Sigma_c|^{\tilde \nu}) \label{eq:scaling_ansatz} 
\end{align}
where the subscript $\pm$ discriminates between above and below $\Sigma_c$ and $c_0$ is some exponent. Imposing that Eq.(\ref{HerschelBulkley}) is recovered for $\Sigma>\Sigma_c$ and $y\rightarrow \infty$ implies $c_0=\beta$. Further imposing that $f(y)$ has a limit independent of $\Sigma-\Sigma_c$ as this quantity goes to zero implies that $\mathcal{F}_{\pm}(x)$ is power-law at vanishing arguments, leading a power-law decay of the fluidity $f(y)\sim y^{-\alpha}$, with an exact relation:
\begin{align}
\alpha = \beta/\tilde\nu.
\label{scaling}
\end{align}
We can thus rewrite Eq.(\ref{eq:scaling_ansatz}) as:
\begin{align}
f(y)=y^{-\beta/\tilde\nu} \mathcal{G}_{\pm}(y |\Sigma-\Sigma_c|^{\tilde \nu}).
\label{00}
\end{align}
where the functions $\mathcal{G}_{\pm}$ impose, when $y\gg \ell(\Sigma)$, a rapid decay of the fluidity to its bulk value,  $f_{b}(\Sigma)=\dot{\gamma(\Sigma)}/\Sigma$. As usual in critical phenomena the scaling form of Eq.(\ref{00}) breaks down at very short distances: when $y$ is of the order of the particle size  the fluidity saturates to a finite value.

\begin{figure*}[htb!]
\centering\includegraphics[width=.9\textwidth]{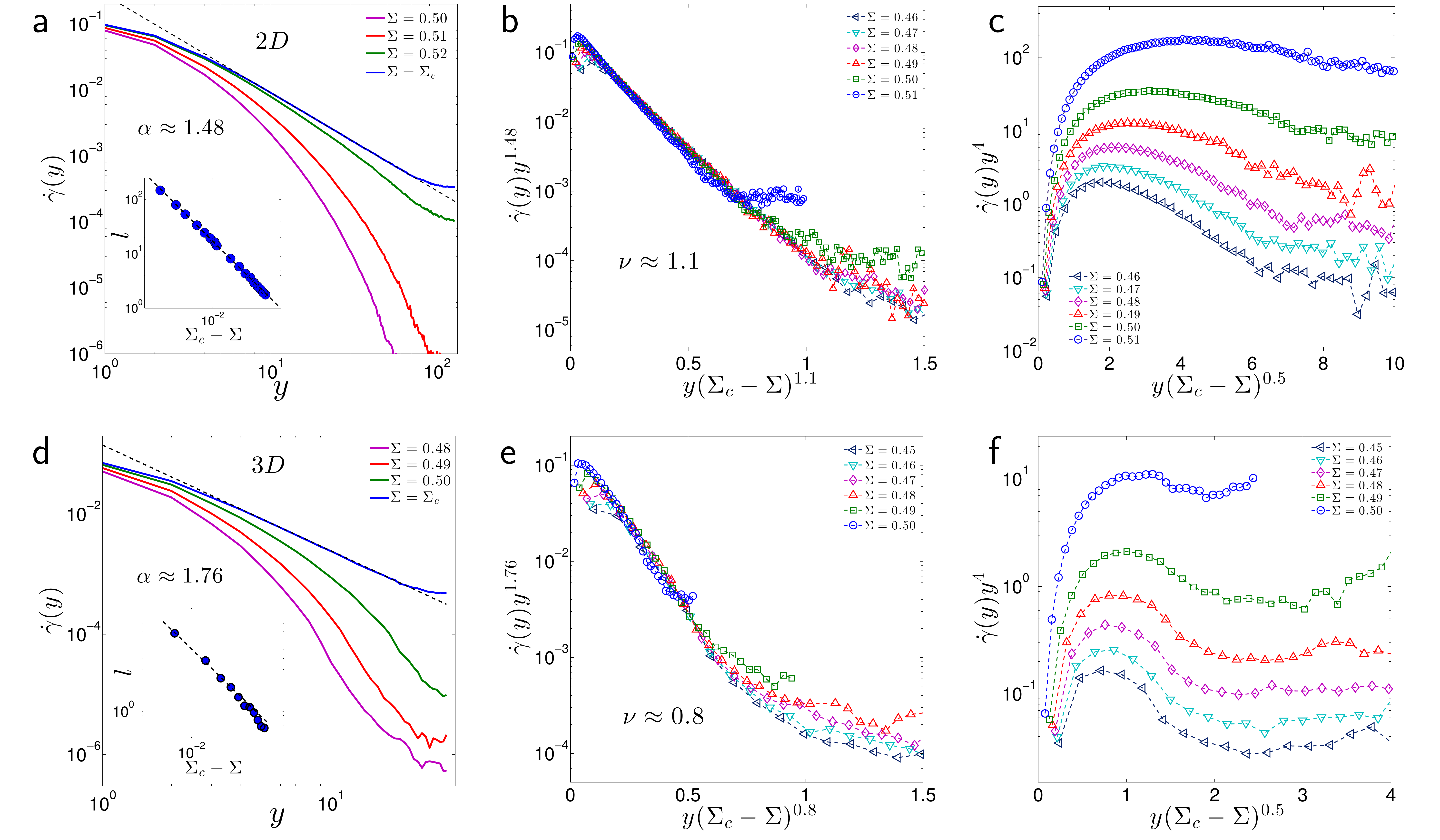} 
   \caption{First column: Strain rate as function of the distance to permenant shear band: $\dot{\gamma}(y)$ {\it v.s.} $y$: (a) $\dot{\gamma}(0)=0.47$, $\Sigma(0)=0.8$ $2D$, $L=256$, (d) $\dot{\gamma}(0)=0.46$, $\Sigma(0)=0.8$ $3D$, $L=64$. Insets are the fits of correlation length (extracted from an exponential fit of $\dot{\gamma}(y) y^\alpha$ shown in (b,e)) as $\ell(\Sigma)\sim(\Sigma_c-\Sigma)^{-\nu}$, where the dashed line has a slope $-1$ in 2D, and $-0.8$ in 3D. Middle column: The corresponding collapses at different stresses, shown in legends, $\dot{\gamma}(y)y^{\alpha}$ {\it v.s.} $y(\Sigma_c-\Sigma)^{\nu}$, where $\nu$ is extracted from the best collapses. Third column: attempt of collapse using the mean field values of exponents: $\alpha_{MF}=4$, $\nu_{MF}=0.5$.}\label{strainratedecay}
\end{figure*}

To  test these results we use  the elasto-plastic model defined in Ref.\cite{Lin14} and specified in  S.I. As shown in Fig.\ref{strainratedecay} (a,d), we indeed find the presence of a power law decay well visible when $\Sigma=\Sigma_c$. To extract exponents, we use the collapse predicted in Eq.(\ref{00}), using for $\Sigma_c$ the values previously obtained in \cite{Lin14} in bulk studies.  The collapse is excellent, shown in Fig.\ref{strainratedecay}(b,e), and we find $\tilde \nu_{2d}=1.1$, $\alpha_{2d}=1.48$ and $\tilde \nu_{3d}=0.8$, $\alpha_{3d}=1.76$. Those results are very close to exponents obtained from bulk studies together with the assumption that $\nu=\tilde \nu$, since we previously measured $\nu_{2d}\approx 1.16$ and $\nu_{3d}\approx 0.72$. Likewise, we had found $\beta_{2d}=1.52$ and $\beta_{3d}=1.38$ implying $\alpha_{2d}\approx 1.38$ and $\alpha_{3d}\approx 1.73$, again in very good agreement  with our measurements. This analysis thus established that $\nu=\tilde \nu$, and henceforth we use the same notation $\nu$ for these quantities. 

By contrast, performing this procedure with mean-field exponents leads to a very poor collapse, as shown in Fig.\ref{strainratedecay}(c,f) using $\alpha_{MF}=\beta_{MF}/\nu_{MF}=4$. This is further evidence that the yielding transition is governed by non-trivial exponents, as predicted \cite{Lin16,Lin14a}.

Finally, it is important to note that Eq.(\ref{standard_fluidity}), often used to fit flow profiles, corresponds to $\alpha=0$ and thus cannot satisfies Eq.(\ref{scaling}). It is thus likely that such fit leads to incorrect extraction for the  length scale characterizing collective effects, which may be responsible for the different behaviors reported. Here we propose the following  differential equation for the flow profile (from which the scaling function $\mathcal{G}_{\pm}$ entering Eq.\ref{00} can be computed):  
\begin{equation}
\ell_0^2 \, \nabla^2 f(\vec x)=C_1 \hbox{sign}(\Sigma_c-\Sigma) |\Sigma_c-\Sigma|^{2\nu}f(\vec x) + C_2f(\vec x)^{\frac{2\nu}{\beta}+1}
\label{nonlocal1}
\end{equation}
where $C_1$, $C_2$ are all constants, and $\ell_0$ sets the microscopic length scale. 
This equation is an extension of  previous proposals in \cite{Bocquet09,Chaudhuri12} that includes non-mean-field exponents and satisfies Eq.\ref{scaling}. In S.I., we show that Eq.(\ref{nonlocal1}) is not exact: there are systematic differences between the measured and predicted flow profile, especially in spatial dimension $d=2$. These deviations however disappear in elasto-plastic models where the interaction kernel between shear transformation is local. This suggests that a purely local differential equation  may be insufficient to describe flow profiles near yielding, for which  elastic interactions between shear transformations are long-range.

\begin{figure}[htb!]
   \includegraphics[trim={1.25cm 1cm 0 1cm},width=.52\textwidth]{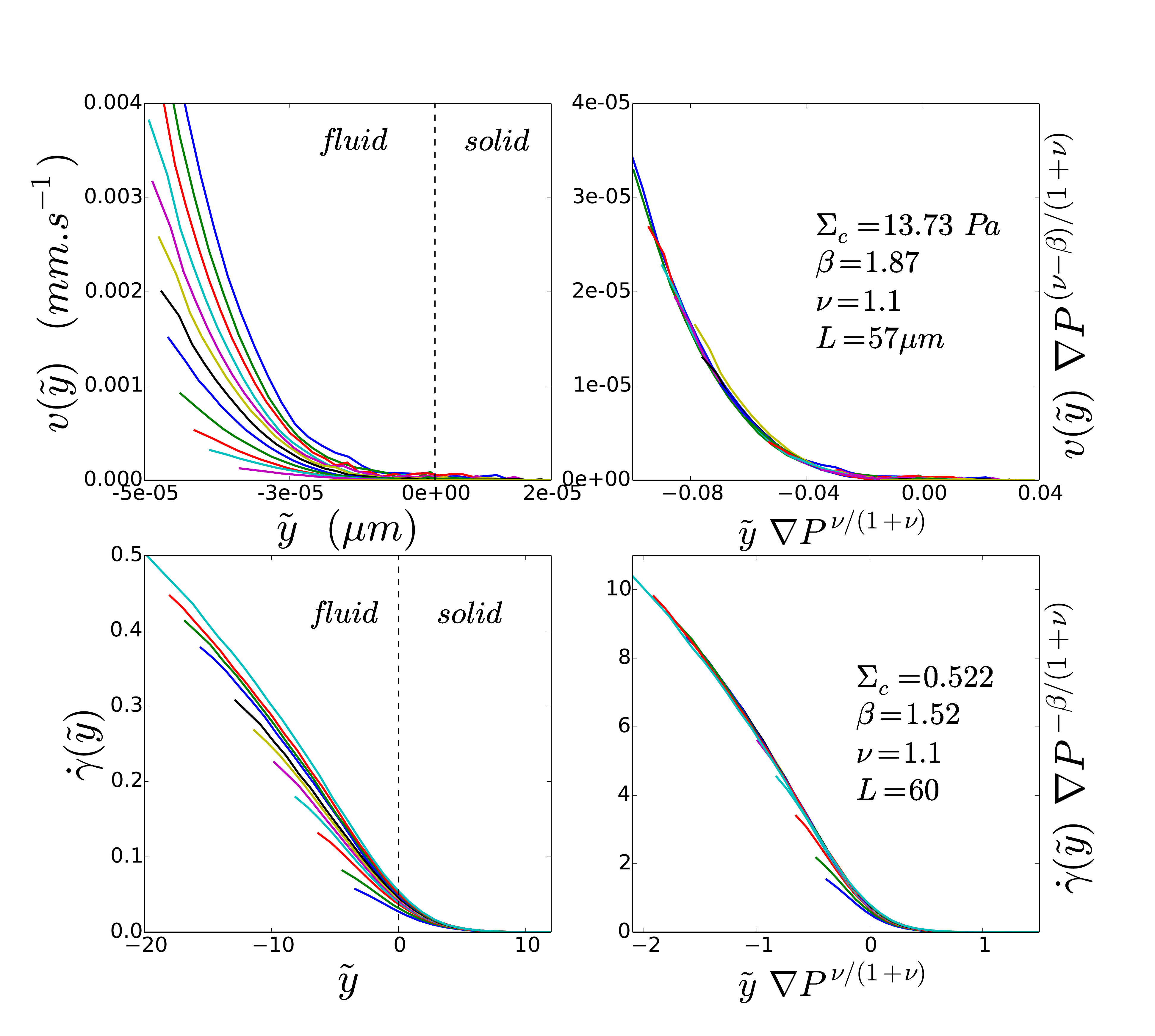} 
    \caption{Collapses of the velocity and activity profiles respectively for experimental datas on Carbopol from \cite{geraud2013confined} (top), and for an 2D elasto-plastic model with periodic Poiseuille geometry (bottom, see SI for a detailed description), following Eq.\ref{eq:scaling_poiseuille}. The left column corresponds to raw profiles, the right column is in rescaled coordinates. In those coordinates, the boundary of the channel is the point of maximal velocity. The steeper the curve, the larger the pressure differential, ranging from $\nabla P = 0.16$ to $0.93$ bar (experiment) and $\nabla P = 10^{-2}$ to $1.7 \cdot 10^{-2}$ (numerics).}
    \label{collapses_poiseuille}
\end{figure}

{\it  Pressure driven Flow}: Our analysis can be extended to pressure driven flows, obtained by applying at the extremities of a channel a pressure gradient. The no-slip condition at the walls enforce the fluid velocity to be zero at the boundaries and induce non uniform shearing stress in the channel. In the simple case of a pipe of width $2L$ with perfect walls, shown in Fig.\ref{illus1} (bottom), the stress drops linearly in $y$ from $\Sigma_w$ to $-\Sigma_w$ \footnote{In realistic systems, we need to account for the roughness and the shape of the wall, but the general theory for this boundary rheology is still missing and we assume the walls rough enough to avoid any stick-and-slip phenomenons \cite{nicolas2013mesoscopic}.}. The value of $\Sigma_w$ is proportional to the pressure gradient along the flow direction, $\nabla P$, the control parameter of this experimental setup. 
It is useful  to introduce $\Delta \Sigma= \Sigma - \Sigma_c$: 
\begin{align}
\Delta \Sigma (\tilde{y}) \equiv \Sigma(\tilde{y}) - \Sigma_c = -\nabla P \tilde{y}
\label{Poiseuille}
\end{align}
with $\tilde{y} = y-y_c$ and $y_c$ defined as $\Sigma(y_c) = \Sigma_c$. 
From Eq.(\ref{Poiseuille}), $\tilde{y}<0$ corresponds to the fluid phase above yielding, while $\tilde{y}>0$ is the solid phase. Using the scaling Ansatz Eq.(\ref{eq:scaling_ansatz}), and replacing $\Sigma-\Sigma_c$ by $\Delta \Sigma (\tilde{y})$ as defined in Eq.(\ref{Poiseuille}), we obtain a scaling form for the fluidity:
\begin{align}
f(\tilde{y}) &= (\nabla P \tilde{y})^{\beta} \mathcal{F}_{\pm}( \tilde y^{1+\nu} \nabla P^{\nu})\\ 
&=\nabla P^{\beta/(1+\nu)} \mathcal{H}_{\pm} \left(\tilde y^{1+\nu} \nabla P^{\nu}\right) \label{eq:scaling_poiseuille}
\end{align}

where $\mathcal{H}$ is a scaling function that can be written in terms of $\mathcal{F}$. Often, experiments rather measure the velocity $v(\tilde{y})$. Assuming the scaling hypothesis remains valid over the whole channel width and using the approximation $\dot{\gamma} (\tilde{y}) \simeq f(\tilde{y}) \Sigma_c$, the scaling form for $v(\tilde{y},\nabla P)$ can be obtained by integration of Eq.\ref{eq:scaling_poiseuille}, leading to:
\begin{align}
v(\tilde{y}) = \nabla P^{(\beta-\nu)/(1+\nu)} \mathcal{I}_{\pm} ( \tilde y^{1+\nu} \nabla P^{\nu})
\end{align}
In Fig.\ref{collapses_poiseuille}, we test such collapses both experimental datas of Carbopol pushed through a channel \cite{geraud2013confined}, and from the previously introduced elasto-plastic model in a microchannel geometry (see S.I. for details). The collapse  works very well, again supporting the validity of our approach. 


Finally, it has been observed \cite{nicolas2013spatial,PhysRevLett.109.036001} that the finite size effects on the yield stress are much more noticeable in pressure driven than in shear driven flows. This also is readily explained by scaling. 
Let us first recall that the yield stress of a finite system size under homogeneous stress is shifted by \cite{henkel2008non}:
\begin{align}
\Sigma_c(L) =\Sigma_c(\infty) + a L^{-1/\nu}
\end{align}
(This is indeed an efficient way to extract $\nu$ \cite{Lin14,Salerno13}). To tackle the specific channel geometry associated with pressure driven flow, following \cite{PhysRevLett.109.036001} we define the analogous quantity $\Sigma_{stop}$ as the value at the wall stress $\Sigma_w$ below which flow stops. We note $\Delta \Sigma_s = \Sigma_{stop} - \Sigma_c$ and aim to estimate its dependence with $L$.

 \begin{figure}[hbt!]
   \includegraphics[width=.45\textwidth]{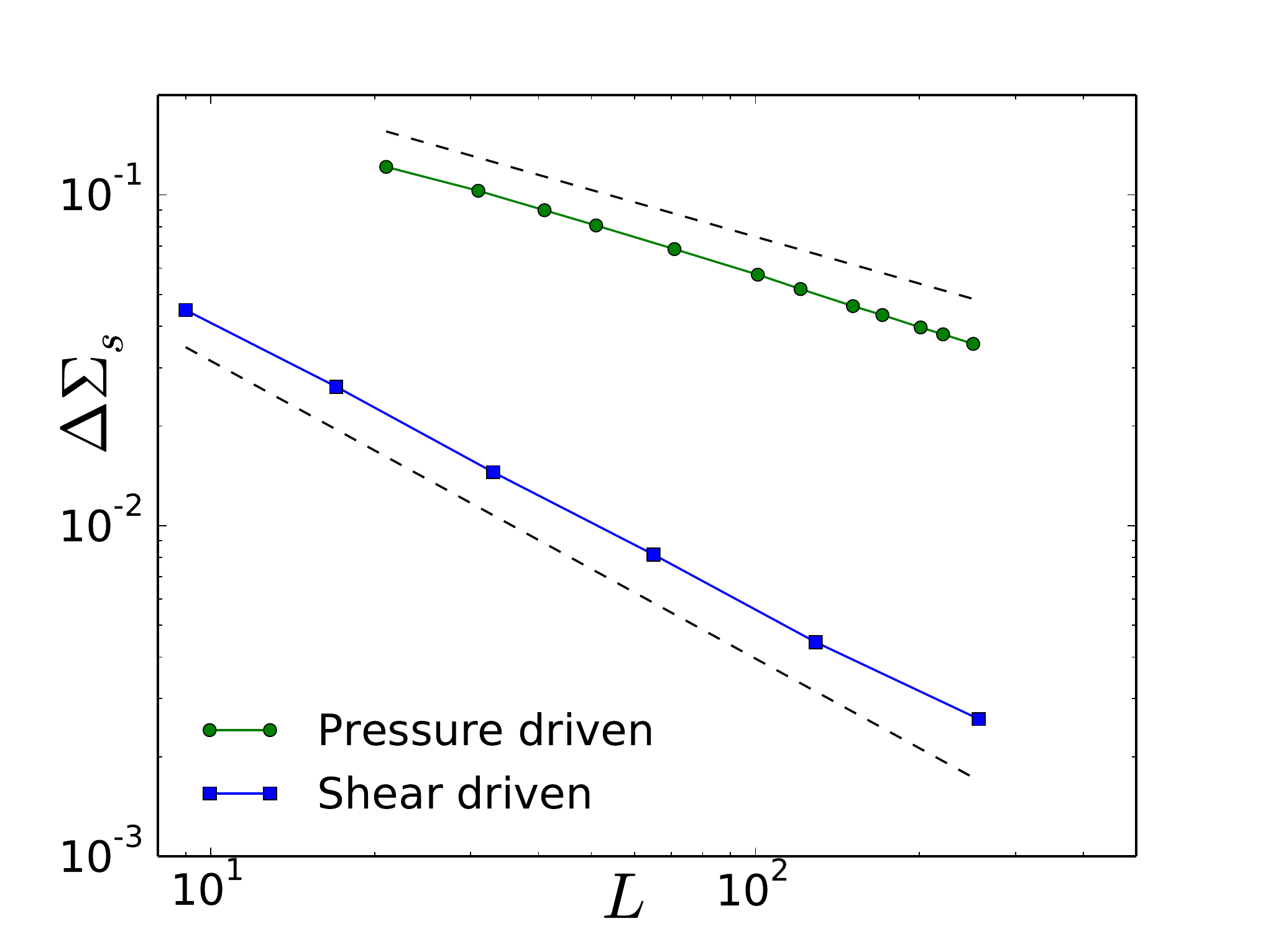} 
    \caption{Scaling analysis of the dependence $\Delta \Sigma_s = \Sigma_c(L)-\Sigma_c$ (squared symbols for shear driven flow) or $\Delta \Sigma_s=\Sigma_{stop}(L)-\Sigma_c$ (circles symbols for pressure driven flow) as a function of $L$ ($\Sigma_c = 0.522$). The dashed black lines are power-law of exponents $-1/\nu \simeq -0.9$ and $-1/(1+\nu) \simeq -0.47$, both consistent with $\nu \simeq 1.1$ for 2D elasto-plastic simulations.}
    \label{finite_effects}
\end{figure}

Let us assume a scenario in which, close to the walls where $\Sigma_w > \Sigma_c$, the flow sets up over a width $l$. We would expect this active band to invade the whole fluid phase $\tilde{y}<0$, so
\begin{align}
l \sim \frac{\Delta \Sigma_s}{\Sigma_w} L\label{lonL}
\end{align}
By definition, $\Sigma_{stop}$ corresponds to the critical stress needed to make a strip of size $l$ flow, $\Sigma_{stop} \sim \Sigma_c + l^{-1/\nu}$. Thus, using $l \sim \Delta \Sigma_s^{- \nu}$ in Eq.(\ref{lonL}) leads to:
\begin{align}
\Delta \Sigma_s \sim \left(\frac{L}{\Sigma_w}\right)^{-1/(1+\nu)} \approx \left(\frac{L}{\Sigma_c}\right)^{-1/(1+\nu)}
\label{band_width}
\end{align}
Such decay, slower than the decay in the shear driven flow, compares well with numerical simulations, as shown in Fig.\ref{finite_effects}. This purely geometrical effect is particularly seen in micro-channels \cite{PhysRevLett.109.036001}.

To summarize, we predict that for $\Sigma_w \simeq \Sigma_{stop}$, the flow in the channel is divided into flowing bands of width $l \sim L^{\nu/(\nu+1)}$ and the rest made of a solid phase, with a (stretched) decaying activity. This picture, quantified here, is referred as \textit{shear localization} \cite{ovarlez2009phenomenology}. 



\hfill

{\it Conclusion:} 
We have provided a scaling description of the yielding transition that unifies both bulk and non-local effects.
It is based on a single length scale $\xi$ characterizing both non-local effects and dynamical correlations in the flowing phase. As predicted \cite{Lin14,Lin14a,Lin16}, the associated exponent $\nu$ is non-mean field, as we clearly confirmed in our numerics. In our description, creep is thus characterized by a diverging length scale as $\Sigma\rightarrow \Sigma_c^-$. This is to be contrasted to quasi-static loading protocols, for which it can be shown \cite{Muller14,Lin15, Ispanovity14} that plasticity can only occur via  system-spanning avalanches ($\xi=\infty$) for all $\Sigma<\Sigma_c$, in agreement with experimental observations of large spatial correlations of plastic zones  \cite{Lebouil2014,Gimbert2013}.

Practically, our approach indicates that collapsing the strain rate profile is the best method to extract bulk properties from non-local effects. Such method will likely reduce the scatter in the experimental estimation of $\nu$, often obtained from exponential fits. It may also allow one to resolve outstanding questions, including the connection between dense granular flows where particles are essentially hard \cite{kamrin2012nonlocal,bouzid2015non}, and the yielding transition of ``soft" particles studies here.

We would like to thank Lyderic Bocquet, G\'eraud Baudoin, Bruno Andreotti and Medhi Bouzid for sharing their datas and  discussions, and E. DeGiuli and E. Lerner for discussions. M.W. thanks the Swiss National Science Foundation for support under Grant No. 200021-165509 and the Simons Collaborative Grant ``Cracking the glass problem".
 
\bibliography{Wyartbibnew}

\newpage
\appendix

\section{Appendix A: Numerical Setup}

We use the well-studied mesoscopic elasto-plastic model to test the above scaling description both in two and three dimensions. We consider square (d=2) and cubic (d=3) lattices with a linear system size $L$, the total number of sites is $N=L^d$. Each lattice point $i$ can be viewed as the coarse grained description of the group of particles in a transformation zone. It is characterized by a scalar stress $\sigma_i$, an instability threshold $\sigma^{th}_i$
 and the plastic strain  $\gamma^{pl}_i$. For simplicity we set $\sigma^{th}_i=\sigma^{th}=1$. A site is unstable if $\sigma_i>1$. At each time step one of the unstable sites is randomly selected and undergoes a plastic deformation that mimics the re-organization of the STZ:
\begin{itemize}
\item we compute $\delta x_i=\sigma_i +\epsilon$ with $\epsilon \in (-0.1,0.1)$,
\item the plastic strain increases locally $\gamma^{pl}_i \to \gamma^{pl}_i +\delta x_i$
\item the local stress of the STZ is re-organized $$
 \sigma_j \to \sigma_j+ {\cal G}( \vec r_{ij})\delta x_i$$
\end{itemize}
here $\vec r_{ij}$ is the vector that joins the site $i$ with $j$. The redistribution kernel has an Eshelby form which in the infinite d=2 system writes as ${\cal G}(\vec r)\propto \cos (4\phi)/r^2$ with  $\phi$  the angle between the shear direction and $\vec{r_{ij}}$. At finite size, periodic boundary conditions are implemented using
its Fourier representation, $G(k_x,k_y) \propto \frac{k_x^2k_y^2}{k^4}$, and the
discrete wave vectors, $k_x=\frac{2 \pi n_x}{L}$, $k_y=\frac{2 \pi n_y}{L}$.  The proportionality constant is fixed by the condition
${\cal G}(0)=-1$. Note that the stress conservation implies that $\sum_{x} G(x,y)=\sum_{y}G(x,y)=0$. The d=3 can be studied using following the same ideas. The yielding transition of this model  has been studied in past works in the bulk and for homogeneous stress \cite{Lin14}. 

\subsubsection{Shear driven flow with homogeneous stress}

To create a permanent shear band, we set $\Sigma(y=0)=0.8>\Sigma_c(0.522)$ and the bulk stress as constant below the yield stress, $\Sigma(y>0)=\Sigma<\Sigma_c$. The strain rate profile $\dot{\gamma}(y)$ is computed as the fraction of active site at distance $y$ per unit time. Our results are stored when the system has reached the steady state after $20N$ local events.

\subsubsection{Pressure driven flow with heterogeneous stress}

The presence of a wall modifies significantly the Eschelby kernel, as one has to take into account the vanishing displacement of the elastic matrix at the boundaries. In that case, the new kernel can be computed exactly \cite{nicolas2013mesoscopic,picard2004elastic}. Yet its expression is rather complicated and does not lend itself easily to numerical computations. A possible workaround is to use a geometry called the Periodic Poiseuille setup, in which the system is divided into two subsystems. Applying a pressure in the opposite direction in each of the two domains and adding periodic boundaries conditions ensures by symmetry that the fluid velocity vanishes at the border of each subsystem. Such setup approximates very well the true pressure driven flows, and was shown \cite{backer2005poiseuille} to be suited for studying rheological properties in simulations. In the context of soft amorphous rheology, it has already been employed for molecular dynamics by the authors of \cite{PhysRevLett.109.036001}. 

The profile of stress for a channel of half-size $L$ (corresponding to a Periodic Poiseuille setup of total size $4L$) with transverse coordinate $y$ is given Eq.(\ref{Poiseuille}). 

%
%
%
%

\section{Appendix B: Scaling functions from fluidity theory}

In this Appendix, we present fitting results obtained from the fluidity equations Eq.\ref{standard_fluidity} and Eq.\ref{nonlocal1} of the main text, very close to yielding threshold. We solve numerically those equations, for both the set of exponents $(\beta,\nu)$ as predicted by the mean-field models, and the ones obtained numerically in finite dimension, as detailed in Appendix A. 

We test those fits on three models: the 2D and 3D elasto-plastic model with the long range Eschelby kernel, as described in Appendix A, and an additional model, with a local kernel, given Table \ref{local}. This model approximates the pattern of correlations between STZ, without exhibiting their long-range decay. Its yielding threshold and the set of exponents have been again determined from simulations in the bulk: $\Sigma_c = 0.598$, $\beta = 0.7 $ and $\nu = 0.84$. 

\begin{figure*}[hbt!]
   \includegraphics[width=1.0\textwidth]{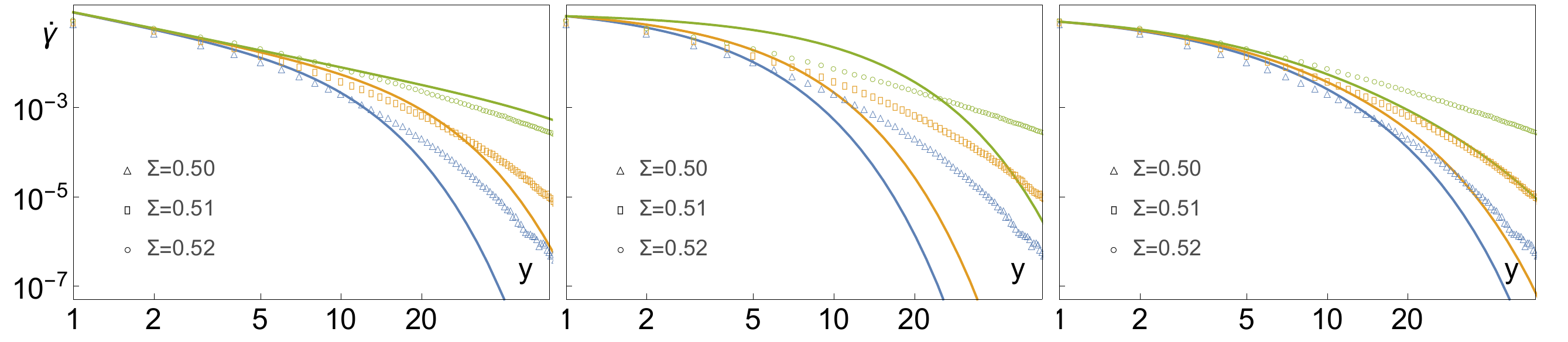} 
    \caption{\textbf{The 2D model with Eschelby kernel}: best fits of the data of Fig 2 of the main text (2D Eshelby kernel) with the fluidity models. From left to right: fit obtained from the generalized mean-field model Eq.\ref{nonlocal1} with $\ell_{0}^2=0.15$; fit obtained from Eq.\ref{standard_fluidity} with $\ell_{0}^2=0.85$; fit obtained from the standard mean-field model Eq.\ref{nonlocal1} with $\ell_{0}^2=0.35$. The bulk exponents were fixed to $\beta_{2d}=1.52$, $\nu_{2d}=1.2$ for the generalized mean-field model, and $\beta_{MF}=2.0$, $\nu_{MF}=0.5$ for the standard mean-field model. The system size is $L=128$. In Eq.\ref{nonlocal1}, we normalize $C_2=1$ and extract $C_1=6.8$ from the HB relation.}
    \label{fit2D}
\end{figure*}

\begin{figure*}[hbt!]
 \includegraphics[width=1.0\textwidth]{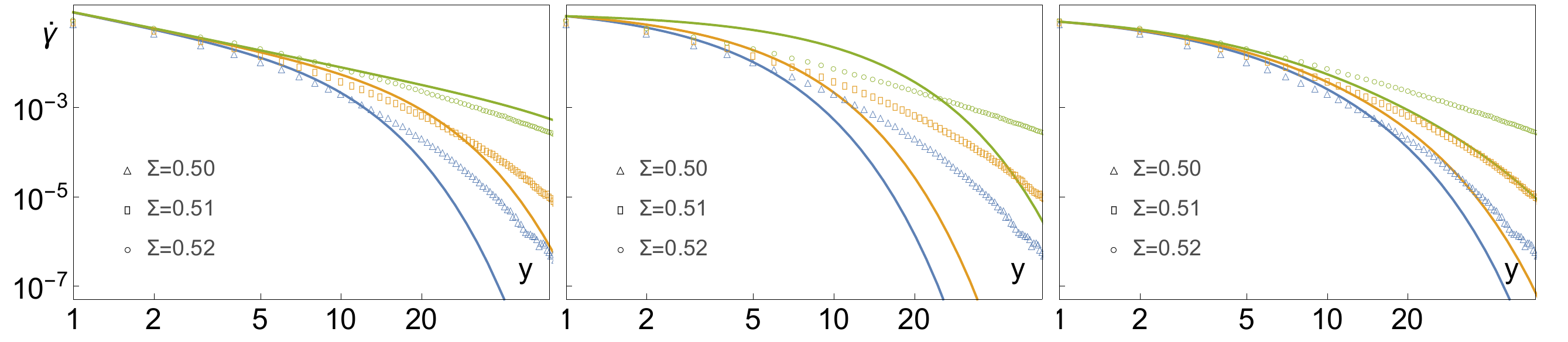} 
    \caption{\textbf{The 3D model with Eschelby kernel}: best fits of the data of Fig 2 of the main text (3D Eshelby kernel) with the fluidity models. From left to right: fit obtained from the generalized mean-field model Eq.\ref{nonlocal1} with $\ell_{0}^2=0.11$; fit obtained from Eq.\ref{standard_fluidity} with $\ell_{0}^2= 0.35$; fit obtained from the standard mean-field model Eq.\ref{nonlocal1} with $\ell_{0}^2=0.58$. The bulk exponents were fixed to $\beta_{3d}=1.4$, $\nu_{3d}=0.8$ for the generalized mean-field model, and $\beta_{MF}=2.0$, $\nu_{MF}=0.5$ for the standard mean-field model. The system size is $L=32$. In Eq.\ref{nonlocal1}, we normalize $C_2=1$ and extract $C_1=3.4$ from the HB relation.}
    \label{fit3D}
\end{figure*}

\begin{figure*}[hbt!]
   \includegraphics[width=1.0\textwidth]{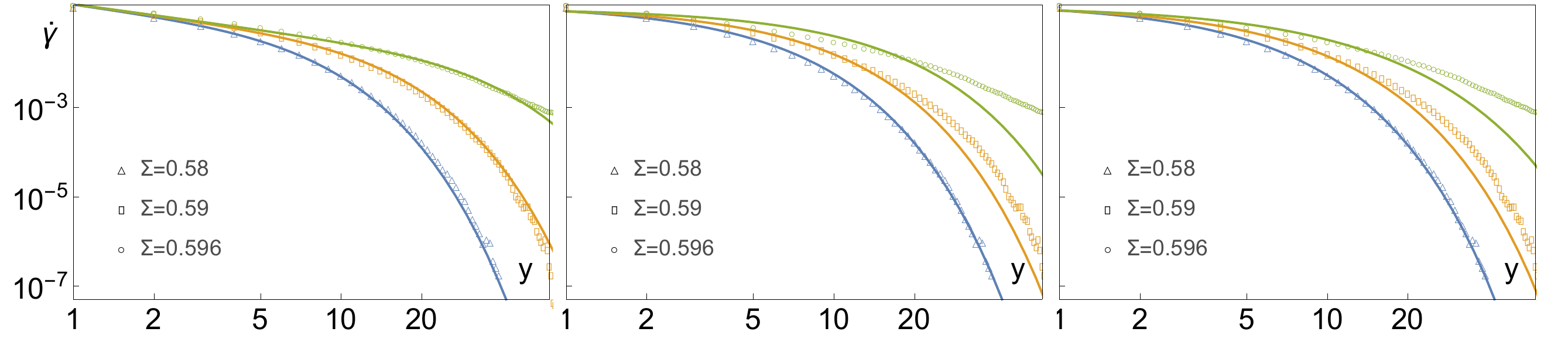} 
    \caption{\textbf{The 2D model with local kernel}: best fits of the data for the 2D local kernel with the fluidity models. From left to right: fit obtained from the generalized mean-field model Eq.\ref{nonlocal1} with $\ell_{0}^2=0.33$; fit obtained from Eq.\ref{standard_fluidity} with $\ell_{0}^2= 0.9$; fit obtained from the standard mean-field model Eq.\ref{nonlocal1} with $\ell_{0}^2=1.0$. The bulk exponents were fixed to $\beta_{local}=0.7$, $\nu_{local}=0.84$ for the generalized mean-field model, and $\beta_{MF}=2.0$, $\nu_{MF}=0.5$ for the standard mean-field model. The system size is $L=128$. In Eq.\ref{nonlocal1}, we normalize $C_2=1$ and extract $C_1=5.1$ from the HB relation.}
    \label{fitlocal}
\end{figure*}

\begin{table}[!htb]
\caption{\small{The local kernel in two dimensions}}\label{local}
\begin{tabular}{ c | c | c  }                  
 $ -1/4$       &  $ 1/2$  &  $ -1/4$     \quad       \\[5pt] \hline
  $ 1/2$       &  $-1$      &  $ 1/2$         \\[5pt] \hline
  $ -1/4$      &  $ 1/2$  &  $ -1/4$    \\[5pt]
\end{tabular}
\end{table}
The parameter $\ell_0$ given in the captions of Fig.\ref{fit2D},\ref{fit3D},\ref{fitlocal} is a \textit{model-dependent} fitting parameter, accounting for the non-local effects. In Eq.\ref{nonlocal1}, $\ell_0$ simply appears in front of the Laplacian. In Eq.\ref{standard_fluidity}, we make the standard choice $\ell(\Sigma)^2 = \ell_0^2 /|\Sigma-\Sigma_c|$ \cite{nicolas2013spatial}.

As expected, the model Eq.(\ref{standard_fluidity}) as well as the mean-field limit of Eq.(\ref{nonlocal1})  are unable to properly describe the developing power-law of exponent $\alpha$, right at the yielding both for the Eshelby (Fig.~\ref{fit2D} and \ref{fit3D}) and local kernel (Fig.~\ref{fitlocal}). The modified fluidity equation (\ref{nonlocal1})  with the finite dimension exponents works better, but we still find a discrepancy with the numerical data. For the data obtained with the local model instead, the fluidity theory gives a much better result overall. This can be understood from the fact that the Laplacian usually describes well short-range interactions. On the contrary, the decay in $1/r^2$ for $d=2$ for the model with Eschelby kernel can only marginally be approximated by a Laplacian.

\end{document}